\documentclass[aps,prb,twocolumn,reprint,groupedaddress,superscriptaddress,amsmath,amssymb,citeautoscript]{revtex4-2}
\usepackage{romannum}
\usepackage{graphicx}
\usepackage{bm}
\usepackage{cleveref}
\usepackage{color}

\begin{document}

\title{Evolution with Magnetic Field of Discrete Scale Invariant
Supercritical States in Graphene}
\author{Hailong Li}
\affiliation{International Center for Quantum Materials, School of Physics,
Peking University, Beijing 100871}
\author{Haiwen Liu}
\affiliation{Center for Advanced Quantum Studies, Department of Physics,
Beijing Normal University, Beijing 100875}
\author{Robert Joynt}
\email{rjjoynt@wisc.edu}
\affiliation{Kavli Institute of Theoretical Sciences, Chinese Academy of
Sciences, Beijing 100049} 
\affiliation{Department of Physics, University of
Wisconsin-Madison,1150 Univ. Ave., Madison WI 53706 USA}
\author{X. C. Xie}
\email{xcxie@pku.edu.cn}
\affiliation{International Center for Quantum Materials, School of Physics,
Peking University, Beijing 100871} 
\affiliation{Beijing Academy of Quantum
Information Sciences, Beijing 100193, China} 
\affiliation{CAS Center for
Excellence in Topological Quantum Computation, University of Chinese Academy
of Sciences, Beijing 100190, China}
\date{\today }

\begin{abstract}
We investigate the quasi-bound states of a Coulomb impurity in graphene in
the presence of a magnetic field. \ These states exhibit the dramatic and
rather rare property of discrete scale invariance when the Coulomb potential is
supercritical. \ We show using both Wentzel-Kramers-Brillouin (WKB) approximation and numerical
studies that the supercritical states are converted to subcritical states as
the field is increased. \ The local density of states is calculated and it shows direct signatures of discrete scale invariance. \ In a magnetic field, these signatures are gradually destroyed in a systematic way. \ Hence the effect that we propose can be detected via scanning tunneling microscope experiments. \ The range of magnetic field and energy resolution required are compatible with existing experimental setups. \ These experiments can be performed in a single sample by changing the field; they do not involve changing the nuclear charge.
\end{abstract}

\maketitle

\section{Introduction\label{1}}

Since the successful isolation of graphene in 2007~\cite{RN33}, it has been
an ideal two-dimensional system to realize ultra-relativistic phenomena and
quantum electrodynamics~\cite{RN288,RN287,RN286}. \ One of the most striking
and novel effects is the collapse of an atomic wavefunction when the
nuclear charge parameter $Z$ exceeds a critical value $Z_{c}$ ~\cite{RN297}.
\ This was predicted many years ago~for atoms in free space \cite%
{RN300,RN37,RN264}. Unfortunately, the expected $Z_{c}\sim 170$ made it
difficult to observe this phenomenon~\cite{RN300,RN38}. However, 
quasiparticles in graphene are two-dimensional massless Dirac fermions and
their velocity is $v_{F}\approx 10^{6}m/s$, which is two orders of magnitude
smaller than the speed of light~\cite{RN296,RN165}. \ Thus the effective
fine structure constant in graphene becomes $\alpha =Ze^{2}/\kappa \hbar
v_{F}\sim \mathcal{O}(1)>>\alpha _{0}\approx 1/137$, where $\alpha _{0}$ is
the bare fine-structure constant. \ So graphene provides a promising
platform to investigate the supercritical atomic collapse phenomenon. \ For
this reason, Coulomb scattering and atomic collapse in graphene have aroused
a great deal of interest~\cite%
{RN295,RN289,RN291,RN194,RN43,RN193,RN44,RN191,RN261}. \ Due to Klein
tunneling and zero mass in graphene, there are no true bound states. \ For
the subcritical case $\alpha <|m|$, there are not even sharp resonances \cite%
{RN43}. \ By contrast, in the supercritical regime with $\alpha >|m|$, an
infinite series of quasi-bound states exist, and these states show a
dramatic property: discrete scale invariance (DSI), reminiscent of the
Efimov trimer states ~\cite{RN174,RN247,RN191,RN175,RN265}. \ In particular,
the resonances associated with this atomic collapse were observed
successfully in scanning tunneling microscope~(STM) experiments
~\cite{RN44,RN290,RN41,RN200}. \ 

DSI has also been observed in bulk three-dimensional materials by the
detection of a new type of quantum oscillation in the magneto-resistance $%
R\left( B\right) $ ~\cite{RN36}. \ $R\left( B\right) $ is periodic in $\ln
\left( B\right) $ in certain topological Dirac materials under ultrahigh
magnetic fields. \ This arises from the Coulomb impurity quasi-bound states
in the material ~\cite{RN265}.

In graphene, samples are usually intercalation compounds of graphene~%
\cite{RN292} or sheets with charged vacancies~\cite{RN187}. \ In the
graphene experiments cited above it is necessary to prepare different
samples with different values of $\alpha $ to test the existence of these
quasi-bound states~\cite{RN298}. Here, we propose that even in a single
sample with a given supercritical $\alpha $, the magnetic field can induce
qualitative changes in the quasi-bound states. \ This leads to an observable
transition from the supercritical regime to the subcritical one. One can
understand the basic idea using a semiclassical argument: the application of 
$\mathbf{B}$ in the z-direction produces a flux through the electronic
wavefunction, and consequently the angular momentum number $m$ will be
modified into an effective quantum number $m_{eff}$. This means that a
quasi-bound state in the presence of a potential with a fixed effective $%
\alpha $ will lead to a transition from supercritical (DSI) regime to the
subcritical (non-DSI) regime when $B$ is increased. \ $B$ provides a
long-distance cutoff at a distance $l_{B}=\sqrt{\hbar c/eB}$ and converts
supercritical states with effective radius longer than $l_{B}$ into
subcritical states. \ Some states should undergo this conversion 
when B is in the range of a few Tesla.

In the Sec. II, we give analytical results to support the basic
argument. \ In Sec. III we present detailed numerical calculations that can
be compared with experiments. \ Sec. IV is a short summary.

\begin{figure}[th]
\includegraphics[width=\columnwidth]{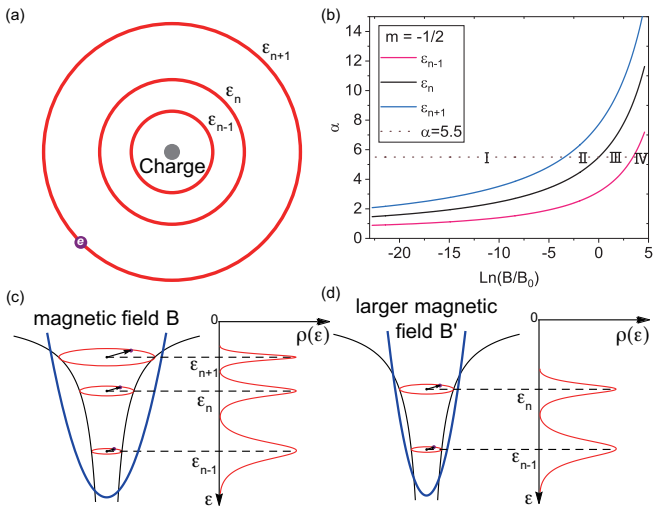}
\caption{(color online) (a) Semiclassical picture of the orbits of the
supercritical states with energies, $\protect\epsilon _{n+1}$, $\protect%
\epsilon _{n}$ and $\protect\epsilon _{n-1}$, with $m=-1/2$, $\protect\alpha %
=5.5$ and zero magnetic field. The radii exhibit the discrete scale
invariance of these states. (b) The transitions of the three states from
supercritical to subcritical. \ $B_{0}$ is related to the short range
cut-off $r_{a}$ with $\protect\sqrt{\hbar c/eB_{0}}=r_{a}$. It shows the
critical $\protect\alpha $ approaches the limit 1/2 as the magnetic field $B$
goes to 0. Moreover, the points of intersection of the dash-dotted line and
the three solid curves give us the critical magnetic field which converts
the supercritical state into a subcritical one with $\protect\alpha =5.5$.
(c) and (d) A schematic picture of the process by which a supercritical
state is converted to a subcritical one as $B$ increases. }
\label{fig:one}
\end{figure}

\section{Theoretical Model\label{2}}

In this section we review the $B=0$ case that can be solved exactly and show
that a kind of Wentzel-Kramers-Brillouin (WKB) approximation gives good
results for finite $B$ in the regime of interest. \ The 2D massless Dirac
equation with a Coulomb potential is

\begin{figure*}[th]
\includegraphics[width=2\columnwidth]{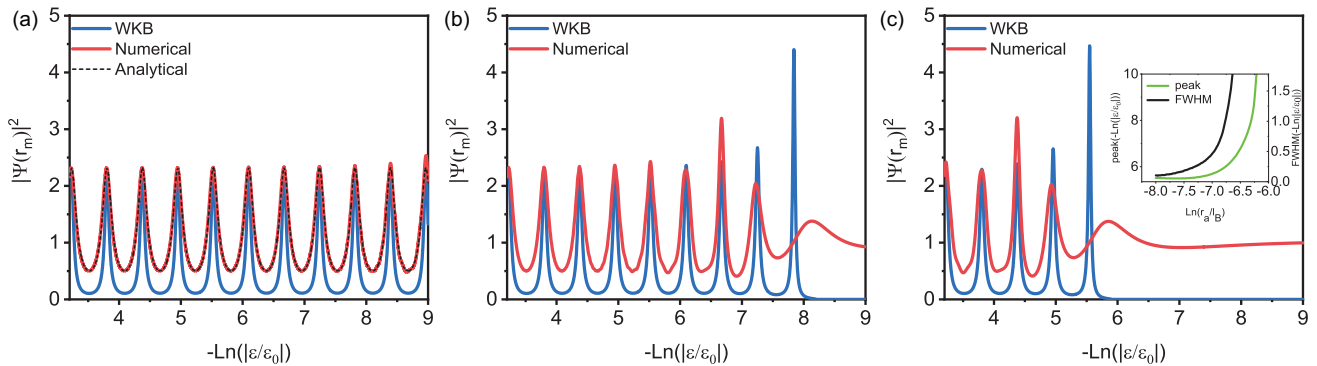}
\caption{(color online) LDOS for the angular momentum channel $m=-1/2$ at a
given radius $r_{m}$. Here $\protect\alpha =5.5$, the dimensionless energy
is plotted on a log scale and $\protect\epsilon _{0}$ is related to the
fermi energy. The pink solid line, blue solid line and the black dashed line
denote the results given by the numerical method, WKB method and analytical
method respectively. As $B$ increases, the supercritical states with smaller binding energies are converted to the subcritical ones earlier than the larger ones. The inset in (c) shows the peak broadening and the binding energy decreasing under increasing magnetic field for a given quasi-bound state. The solutions are calculated under three different magnetic fields: (a) $%
l_{B}=10^6r_{a}$; (b) $l_{B}=5\times10^4r_{a}$; (c) $l_{B}=5\times10^3r_{a}$. Here, $%
r_{a}$ is a cut-off for the Coulomb potential.}
\label{fig:two}
\end{figure*}

\begin{equation}
\frac{1}{\hbar }\left( \bm{\sigma}\cdot \left( \bm{p}+e\bm{A}/c\right) -%
\frac{\alpha }{r}\right) \psi (\bm{r})=\varepsilon \psi (\bm{r}),
\label{eq:one}
\end{equation}%
where $\psi (\bm{r})$ is a two-component wave function, $\bm{\sigma}$
denotes the Pauli matrices, and $\varepsilon =E/\hbar v_{F},$ where $E\ $is
the energy. \ We adopt the symmetric gauge for the magnetic vector
potential, $\bm{A}=\left( -By/2,Bx/2,0\right) $. \ In addition we impose a
short distance cutoff at $r=r_{a}$ where $r$ is the radial coordinate such
that the potential $V\left( \mathbf{r}\right) =Ze^{2}/\kappa \hbar cr_{a}$
for $r<r_{a}.$ \ The region $r_{a}<<r<<l_{B}$ is the scale-invariant region:
when $r$ is in this range, all operators have dimension (length)$^{-1}.$\ 

We firstly review the analytic solution to the $B=0$ case. By introducing
cylindrical coordinates, we can separate the equation and obtain a general
solution in the following form: 
\begin{equation}
\psi (r,\phi )=\frac{u_{2}}{\sqrt{r}}\left( 
\begin{array}{l}
1 \\ 
0%
\end{array}%
\right) e^{i(m-1/2)\phi }+i\frac{u_{1}}{\sqrt{r}}\left( 
\begin{array}{c}
0 \\ 
1%
\end{array}%
\right) e^{i(m+1/2)\phi }\label{eq:four}
\end{equation}%
where $m$ is a half-integer angular momentum quantum number, $u_{1}(r)$ and $%
u_{2}(r)$ are the radial functions. \ For $\alpha <|m|$ or subcritical case,
the solutions do not yield any quasi-bound states, and the only non-trivial
effect is the so-called Coulomb phase shift~\cite{RN259}. \ When $\alpha
>\left\vert m\right\vert $, the supercritical case, a cut-off for the
Coulomb potential at a small $r_{a}$ is necessary for avoiding a divergence~%
\cite{RN293,RN294}. \ We choose a soft cutoff such that $V\left( r\right)
=Ze^{2}/\kappa r_{a}$ for $r\leq r_{a}.$ \ The results are not sensitive to
the details of the cutoff. \ The states are sharp resonances with
quasi-energies~\cite{RN44} 
\begin{equation}
\frac{\varepsilon _{n+1}}{\varepsilon _{n}}\propto e^{-\pi /\sqrt{\alpha
^{2}-m^{2}}},  \label{eq:two}
\end{equation}%
where $n$ is the radial (principal) quantum number. \ The quasi-energies
forms a geometric series. \ We give a schematic picture of the states in
Fig.~\ref{fig:one}a. 

For $B>0$ one can apply the WKB approximation to Eq.~(\ref{eq:one}) to
describe the supercritical quasi-bound states when $\alpha $ exceeds a
certain value that now depends on $B.$ \ This method will also allow us to
understand how the external magnetic field can be thought of as an effective
angular momentum, $m_{eff}$, replacing the angular momentum $m$ of the $B=0$
case. The appropriate radial momentum is 
\begin{equation}
p_{r}^{2}=\left( \frac{E}{\hbar v_{F}}+\frac{\alpha }{r}\right) ^{2}-\left( 
\frac{m}{r}-\frac{r}{2l_{B}^{2}}\right) ^{2}.  \label{eq:three}
\end{equation}%
\ \ The Bohr-Sommerfeld quantization condition is $%
\int_{r_{0}}^{r_{1}}p_{r}dr=n\pi \hbar $, where $r_{0}$ and $r_{1}$ are turning
points of Eq.~(\ref{eq:three}). \ \ The quasi-bound states form in the
classically permitted region and they can tunnel to the outer region via
Klein tunneling. With the WKB method, scattering coefficients can be
obtained analytically and the numerical results of the local density of states
are shown in Fig.~\ref{fig:two}.

The condition for the transition from supercritical to subcritical can be
written in an intuitive form as follows. For the $n$th state, we rewrite $%
r/(2l_{B}^{2})$ as $\bm{B}\cdot\bm{S_n}/\left( 2\Phi _{0}r\right) $ , where $S_{n}$ is the
rms area of the part of the wavefunction of the $n$th supercritical state
that lies inside the potential barrier. $\Phi _{0}=h/(2e)$ denotes the
magnetic flux quantum. So, it is natural to define $\bm{B}\cdot\bm{S_n}/\left( 2\Phi
_{0}\right) $ as a new physical quantity, $N_{B,n}$, which is one half the
number of magnetic flux quanta enclosed by the $n$th supercritical state in
the scale-invariant region. \ Then, we also define the absolute value of the
effective angular momentum quantum number as $\left\vert m_{eff}\right\vert=\left\vert m-N_{B,n}\right\vert$. \ Thus the
Bohr-Sommerfeld condition gives a relation $\alpha \left( B\right) $ that is
plotted in Fig.~\ref{fig:one}b. \ As $B$ increases, the $n$th state makes a
transition from supercritical to subcritical when $\alpha =\left\vert
m_{eff}\right\vert .$ \ Taking the physical value of $\alpha ,$ we get a
critical value $B_{n}$ for which there is a sudden transition from a very
narrow resonance to an extremely broad one. \ We show this for three states
and $\alpha =5.5.$ \ As we shall see, their identifiable contribution to the
LDOS effectively disappears as they successively pass through the
transition.  In Fig 1c all three states are supercritical; Fig. 1d
corresponds to the case where the $n-1$ and $n$ state are still visible but
the $n+1$ state is not. \ 

\section{Numerical Approach\label{3}}

The 2D massless Dirac equation with $B>0$ can be solved numerically. This is
necessary since the WKB approximation is always poor near the turning
points. \ However, the numerics will show that WKB is still reliable as a
semi-quantitative guide. 

We start from the radial equation, Eq.~(\ref{eq:seven}), and use the finite
difference method. The radial equation can be derived by substituting Eq.~(%
\ref{eq:four}) into Eq.~(\ref{eq:one}), which gives 
\begin{equation}
\frac{d}{dr}\left[ 
\begin{array}{c}
u_{1}(r) \\ 
u_{2}(r)%
\end{array}%
\right] =\left[ 
\begin{array}{cc}
-\frac{m}{r}+\frac{r}{2l_{B}^{2}} & \frac{E}{\hbar v_{F}}-V(r) \\ 
-\frac{E}{\hbar v_{F}}+V(r) & \frac{m}{r}-\frac{r}{2l_{B}^{2}}%
\end{array}%
\right] \left[ 
\begin{array}{c}
u_{1}(r) \\ 
u_{2}(r)%
\end{array}%
\right] .  \label{eq:seven}
\end{equation}%

In Eq.~(\ref{eq:seven}), $V(r)$ is the modified Coulomb potential with the
above-mentioned soft cutoff at $r_{a}$. \ Eliminating $u_{2}$ gives the
final approximate radial equation for $u_{1}$ as: 
\begin{equation}
u_{1}^{\prime \prime }(r)=\left[\left( \frac{m}{r}-\frac{r}{2l_{B}^{2}}\right)
^{2}-\left( \frac{E}{\hbar v_{F}}-V(r)\right) ^{2}\right]u_{1}(r),~
\label{eq:eight}
\end{equation}%
after which $u_{2}$ is easily solved.

The following discussion is based on numerical solutions of these equations.

\begin{figure*}[th]
\includegraphics[width=2\columnwidth]{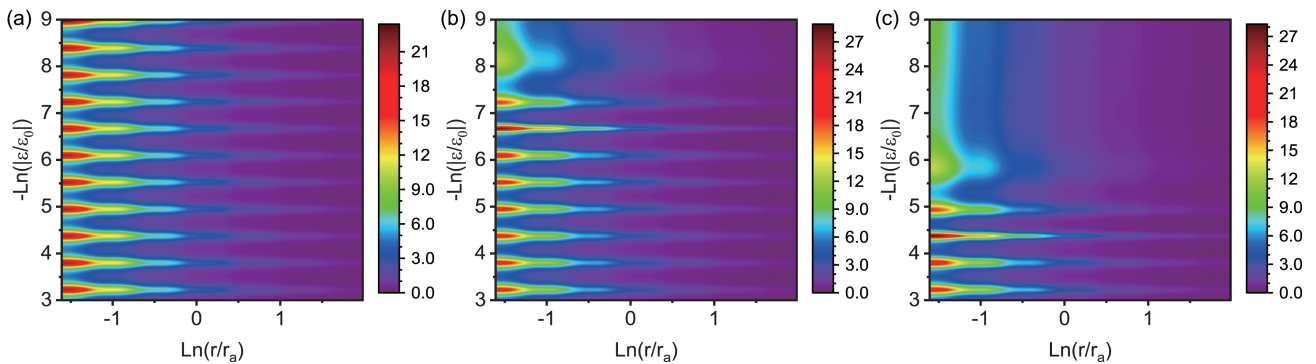}
\caption{(color online) Local density of states for the angular momentum
channel $m=-1/2$ and potential strength $\protect\alpha =5.5.$ \ The LDOS is
plotted as a function of energy and distance away from the charged impurity
for three different magnetic fields on a log scale. Both discrete scale
invariance and the transitions from supercritical to subcritical are
directly reflected in the LDOS. (a) $%
l_{B}=10^6r_{a}$; (b) $l_{B}=5\times10^4r_{a}$; (c) $l_{B}=5\times10^3r_{a}$.}
\label{fig:three}
\end{figure*}

\subsection{Validation of Numerics}

The LDOS for the lowest angular momentum channel $m=-1/2$ obtained by 
three different methods is shown in Fig.~\ref{fig:two}, enabling the reader
to compare the different methods. In Fig.~\ref{fig:two}, we set $\alpha =5.5$%
, and plot the LDOS at an arbitrarily given location $r_{m}$ in logarithmic
scale. In Fig.~\ref{fig:two}a, the magnetic field is so weak that it almost
does not affect the effective angular momentum of the quasi-bound states and
they all remain supercritical. The results given by analytical solution and
numerical method fit each other very well, and the results of WKB
approximation shows similar peak values but slightly different widths. In the
log-scale energy plot shown in Fig.~\ref{fig:two}, peak values behave as
an arithmetic sequence, which reveals the DSI of these quasi-bound states.
For larger $B,$ the analytic method is inapplicable, so only the WKB\ and
finite-difference results are plotted in Fig.~\ref{fig:two}b,c. \ As $B$
increases, the quasi-bound states begin to ``disappear'' -- their width
increases so rapidly that they leave no signature in the LDOS. As an illustration, the inset in Fig.~\ref{fig:two}c shows that the binding energy of a given quasi-bound state decreases and the corresponding full width half maximum(FWHM) becomes wider with the increasing magnetic field. The numerical results agree well with the WKB results, except for the state with smallest binding energy, of which the energy peak shows a minor difference. However, the transition fields for the ``disappearance'' of the quasi-bound state are nearly the same with the two methods. \ These
results confirm the argument of the previous section that these transitions
occur when $\left\vert m_{eff}\right\vert =\alpha .$ 

\subsection{Local Density of States in STM Experiments}

We calculate the LDOS as a function of energy and distance away from the
charged impurity by the numerical method for a few representative values of $%
B$, with the results shown in Fig.~\ref{fig:three}. The three subgraphs are all
plotted on log scale. Larger numbers in the vertical axis refer to smaller binding
energies, while smaller numbers refer to larger binding energies. Fig.~\ref%
{fig:three}a shows a series of supercritical states for small $B$. The
typical periodicity in $\ln \varepsilon $ is evident. \ The overall physical
picture is strikingly confirmed by the appearance of one broad resonance
between the series of narrow resonances and the continuum at small energies.
The larger magnetic field in Fig.~\ref{fig:three}b converts some
supercritical states with smaller binding energies into subcritical ones with the rest of the quasi-bound states still satisfying discrete scale invariance. \ More
supercritical states disappear due to a much larger magnetic field in Fig.~%
\ref{fig:three}c. Therefore, with increasing magnetic field,
the supercritical states are converted to subcritical ones and disappear
from small binding energies to large binding energies. Although the external magnetic field
affects the existence of some quasi-bound states, it does not destroy the
distribution of the rest of the supercritical states and the property of discrete
scale invariance.

The results in Fig.~\ref{fig:three} can be directly checked in STM
experiments under a magnetic field. \ For example, if we draw a vertical line 
at a given radius $r_{m}$ in Fig.~\ref{fig:three} the
oscillations will be visible as the voltage is varied. When the experiment
is repeated at larger $B,$ the oscillations in the LDOS will disappear one
by one.

\begin{figure}[ht!]
\includegraphics[width=\columnwidth]{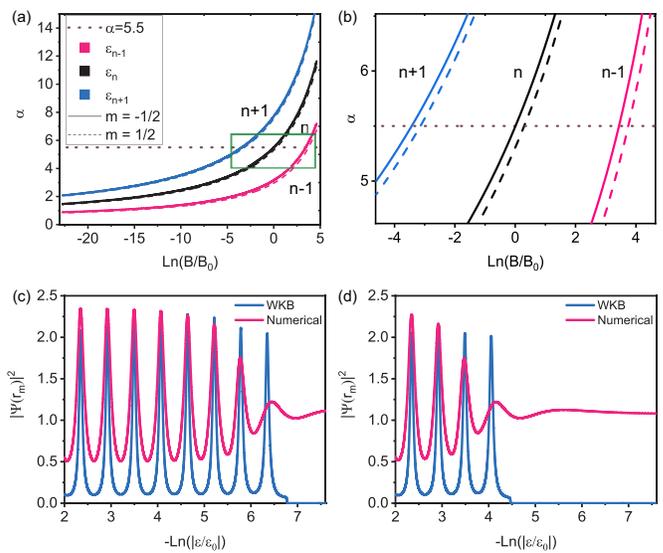}
\caption{(color online) (a) The critical $\protect\alpha$ as a function of
the magnetic field. It shows the critical $\protect\alpha$ approaches the
limit 1/2 as the magnetic field B goes to 0. Here, $B_0$ is defined as $%
\protect\sqrt{\hbar/(eB_0)}=r_a$. For $\protect\alpha=5.5$, $\Delta
Log(B/B_0)$ between $\protect\epsilon_{n+1}$ and $\protect\epsilon_{n}$ is $%
-3.44$ as well as $-3.43$ between $\protect\epsilon_{n}$ and $\protect%
\epsilon_{n-1}$ when $m=-1/2$. As for $m=1/2$, $\Delta Log(B/B_0)$ between $%
\protect\epsilon_{n+1}$ and $\protect\epsilon_{n}$ is $-3.44$, while the one
between $\protect\epsilon_{n}$ and $\protect\epsilon_{n-1}$ is $-3.45$. It
proves that the degeneracy breaking by the magnetic field won't affect the
observation of the supercritical states and the process in which the
supercritical states converts to the subcritical states especially for large 
$\protect\alpha$. (b) the zoom-in plot of the green rectangle in Fig.~%
\protect\ref{fig:four}a. (c) LDOS obtained by numerical method and WKB
method with the magnetic length $l_B=5\times10^4r_a$ and $m=1/2$. (d) LDOS obtained
by numerical method and WKB method with the magnetic length $l_B=5\times10^3r_a$ and 
$m=1/2$.}\label{fig:four}
\end{figure}

\subsection{Multiple Angular Momentum Channels}

In the previous section we focused on a single angular momentum channel,
which produced a relatively simple picture of the LDOS. \ When multiple
channels are present, the situation is more complicated. \ We first note that
this physical system has time-reversal symmetry when $B=0$, which guarantees
the degeneracy between $m$ and $-m$ channels. \ These degenerate quasi-bound states
separately satisfy DSI. The non-zero magnetic field will break the
degeneracy, but\ DSI can not be hidden if there is a near degeneracy. \ 

For further comparison, we calculate the LDOS for the case of $m=1/2$, the
counterpart of $m=-1/2$. In Fig.~\ref{fig:four}c and \ref{fig:four}d, we
compare the different solutions calculated by the numerical method and WKB
method with different magnetic fields as we did in Fig.~\ref{fig:two}. It
behaves the same as in the case of $m=-1/2$. The log-scale energy levels
behave as an arithmetic sequence and the log-scale energy difference is
nearly the same as that of $m=-1/2$, and the presence of DSI is clear. Once
again, as $B$ increases, the supercritical states with smaller binding energies are converted
to the subcritical states earlier than the larger ones.

For an arbitrary magnetic field, we use the Bohr-Sommerfeld quantization
condition as in Fig.~\ref{fig:one}b to calculate the critical $\alpha $ as a
function of the magnetic field for given supercritical states. Here, we only
consider the case when the Landau level spacing is smaller than the Coulomb
attraction energy. The comparison of $m=-1/2$ and $m=1/2$ is shown in Fig.~%
\ref{fig:four}a. Here, the solid lines and the dashed lined represent $m=-1/2
$ and $m=1/2$ respectively. The critical $\alpha $ approaches the limit $1/2$
as $B$ goes to $0$. As we have discussed in Sect.~\ref{1}, these critical
magnetic fields also satisfy DSI for both cases.\ As shown in Fig.~\ref{fig:four}b, with enlarging the magnetic field along the horizontal dash-dotted line, the n-th supercritical state becomes subcritical when crossing the line labelled by n.

We have focused on the lowest angular momentum channels, since the higher ones are less likely to show up in experiments. The degeneracy breaking resulting
from the external magnetic field does not cause significant differences
between $m$ and $-m$, especially for large $\alpha $. Guaranteed by these
aforementioned conditions, the evolution process of the supercritical states
can be observed in experiments by STM. 
From previous experiments~\cite{RN36}, the value $\alpha =5.5$  seems
the most reasonable. Assuming a cutoff $r_a = 0.5 nm$, we can 
estimate that in the range from $B=10^{-2} T$ to $B = 2T$, three or four 
quasi-bound states will disappear.  This is a convenient range for
many experiments.  Their energies are in the range of 10 meV, so very high energy resolution is not required.

\section{Summary\label{4}}

We have investigated the LDOS of quasi-bound states in graphene with a
supercritical Coulomb potential. These supercritical states
exhibit a dramatic property: discrete scale invariance. We propose that the
external magnetic field can change the effective angular momentum for each
definite supercritical state and convert it to a subcritical one. We show
the local density of states for a given angular momentum channel, an
observable quantity, reveals directly that these supercritical states vanish
one by one starting from small binding energies and proceeding to large binding energies as the
field increases. \ Moreover, the critical magnetic field values also
satisfies the property of discrete scale invariance. Our proposal can be
verified by LDOS measurement in STM experiments.

\begin{center}
\textbf{Acknowledgements}
\end{center}

We thank Ziqiang Wang and Qing-Feng Sun for valuable discussions. This work
was financially supported by the National Basic Research Program of China
(Grants No. 2017YFA0303301, No. 2015CB921102), the National Natural Science
Foundation of China (Grants No. 11674028, No. 11534001, No. 11504008), and
the Fundamental Research Funds for the Central Universities.


\begin{thebibliography}{34}%
    \makeatletter
    \providecommand \@ifxundefined [1]{%
     \@ifx{#1\undefined}
    }%
    \providecommand \@ifnum [1]{%
     \ifnum #1\expandafter \@firstoftwo
     \else \expandafter \@secondoftwo
     \fi
    }%
    \providecommand \@ifx [1]{%
     \ifx #1\expandafter \@firstoftwo
     \else \expandafter \@secondoftwo
     \fi
    }%
    \providecommand \natexlab [1]{#1}%
    \providecommand \enquote  [1]{``#1''}%
    \providecommand \bibnamefont  [1]{#1}%
    \providecommand \bibfnamefont [1]{#1}%
    \providecommand \citenamefont [1]{#1}%
    \providecommand \href@noop [0]{\@secondoftwo}%
    \providecommand \href [0]{\begingroup \@sanitize@url \@href}%
    \providecommand \@href[1]{\@@startlink{#1}\@@href}%
    \providecommand \@@href[1]{\endgroup#1\@@endlink}%
    \providecommand \@sanitize@url [0]{\catcode `\\12\catcode `\$12\catcode
      `\&12\catcode `\#12\catcode `\^12\catcode `\_12\catcode `\%12\relax}%
    \providecommand \@@startlink[1]{}%
    \providecommand \@@endlink[0]{}%
    \providecommand \url  [0]{\begingroup\@sanitize@url \@url }%
    \providecommand \@url [1]{\endgroup\@href {#1}{\urlprefix }}%
    \providecommand \urlprefix  [0]{URL }%
    \providecommand \Eprint [0]{\href }%
    \providecommand \doibase [0]{https://doi.org/}%
    \providecommand \selectlanguage [0]{\@gobble}%
    \providecommand \bibinfo  [0]{\@secondoftwo}%
    \providecommand \bibfield  [0]{\@secondoftwo}%
    \providecommand \translation [1]{[#1]}%
    \providecommand \BibitemOpen [0]{}%
    \providecommand \bibitemStop [0]{}%
    \providecommand \bibitemNoStop [0]{.\EOS\space}%
    \providecommand \EOS [0]{\spacefactor3000\relax}%
    \providecommand \BibitemShut  [1]{\csname bibitem#1\endcsname}%
    \let\auto@bib@innerbib\@empty
    \bibitem [{\citenamefont {Geim}\ and\ \citenamefont {Novoselov}(2007)}]{RN33}%
      \BibitemOpen
      \bibfield  {author} {\bibinfo {author} {\bibfnamefont {A.~K.}\ \bibnamefont
      {Geim}}\ and\ \bibinfo {author} {\bibfnamefont {K.~S.}\ \bibnamefont
      {Novoselov}},\ }\href {https://doi.org/10.1038/nmat1849} {\bibfield
      {journal} {\bibinfo  {journal} {Nat. Mater.}\ }\textbf {\bibinfo {volume}
      {6}},\ \bibinfo {pages} {183} (\bibinfo {year} {2007})}\BibitemShut {NoStop}%
    \bibitem [{\citenamefont {Katsnelson}\ and\ \citenamefont
      {Novoselov}(2007)}]{RN288}%
      \BibitemOpen
      \bibfield  {author} {\bibinfo {author} {\bibfnamefont {M.~I.}\ \bibnamefont
      {Katsnelson}}\ and\ \bibinfo {author} {\bibfnamefont {K.~S.}\ \bibnamefont
      {Novoselov}},\ }\href
      {https://doi.org/https://doi.org/10.1016/j.ssc.2007.02.043} {\bibfield
      {journal} {\bibinfo  {journal} {Solid State Commun.}\ }\textbf {\bibinfo
      {volume} {143}},\ \bibinfo {pages} {3} (\bibinfo {year} {2007})}\BibitemShut
      {NoStop}%
    \bibitem [{\citenamefont {Katsnelson}\ \emph {et~al.}(2006)\citenamefont
      {Katsnelson}, \citenamefont {Novoselov},\ and\ \citenamefont {Geim}}]{RN287}%
      \BibitemOpen
      \bibfield  {author} {\bibinfo {author} {\bibfnamefont {M.~I.}\ \bibnamefont
      {Katsnelson}}, \bibinfo {author} {\bibfnamefont {K.~S.}\ \bibnamefont
      {Novoselov}},\ and\ \bibinfo {author} {\bibfnamefont {A.~K.}\ \bibnamefont
      {Geim}},\ }\href {https://doi.org/10.1038/nphys384
      https://www.nature.com/articles/nphys384#supplementary-information}
      {\bibfield  {journal} {\bibinfo  {journal} {Nat. Phys.}\ }\textbf {\bibinfo
      {volume} {2}},\ \bibinfo {pages} {620} (\bibinfo {year} {2006})}\BibitemShut
      {NoStop}%
    \bibitem [{\citenamefont {González}\ \emph {et~al.}(1994)\citenamefont
      {González}, \citenamefont {Guinea},\ and\ \citenamefont
      {Vozmediano}}]{RN286}%
      \BibitemOpen
      \bibfield  {author} {\bibinfo {author} {\bibfnamefont {J.}~\bibnamefont
      {González}}, \bibinfo {author} {\bibfnamefont {F.}~\bibnamefont {Guinea}},\
      and\ \bibinfo {author} {\bibfnamefont {M.~A.~H.}\ \bibnamefont
      {Vozmediano}},\ }\href
      {https://doi.org/https://doi.org/10.1016/0550-3213(94)90410-3} {\bibfield
      {journal} {\bibinfo  {journal} {Nucl. Phys. B}\ }\textbf {\bibinfo {volume}
      {424}},\ \bibinfo {pages} {595} (\bibinfo {year} {1994})}\BibitemShut
      {NoStop}%
    \bibitem [{\citenamefont {Boyer}(2004)}]{RN297}%
      \BibitemOpen
      \bibfield  {author} {\bibinfo {author} {\bibfnamefont {T.~H.}\ \bibnamefont
      {Boyer}},\ }\href {https://doi.org/10.1119/1.1737396} {\bibfield  {journal}
      {\bibinfo  {journal} {Am. J. Phys.}\ }\textbf {\bibinfo {volume} {72}},\
      \bibinfo {pages} {992} (\bibinfo {year} {2004})}\BibitemShut {NoStop}%
    \bibitem [{\citenamefont {Greiner}\ \emph {et~al.}(1985)\citenamefont
      {Greiner}, \citenamefont {Müller},\ and\ \citenamefont {Rafelski}}]{RN300}%
      \BibitemOpen
      \bibfield  {author} {\bibinfo {author} {\bibfnamefont {W.}~\bibnamefont
      {Greiner}}, \bibinfo {author} {\bibfnamefont {B.}~\bibnamefont {Müller}},\
      and\ \bibinfo {author} {\bibfnamefont {J.}~\bibnamefont {Rafelski}},\
      }\href@noop {} {\emph {\bibinfo {title} {Quantum electrodynamics of strong
      fields: with an introduction into modern relativistic quantum mechanics}}}\
      (\bibinfo  {publisher} {Springer},\ \bibinfo {year} {1985})\BibitemShut
      {NoStop}%
    \bibitem [{\citenamefont {Zeldovich}\ and\ \citenamefont {Popov}(1972)}]{RN37}%
      \BibitemOpen
      \bibfield  {author} {\bibinfo {author} {\bibfnamefont {Y.~B.}\ \bibnamefont
      {Zeldovich}}\ and\ \bibinfo {author} {\bibfnamefont {V.~S.}\ \bibnamefont
      {Popov}},\ }\href {https://doi.org/DOI 10.1070/PU1972v014n06ABEH004735}
      {\bibfield  {journal} {\bibinfo  {journal} {Sov. Phys. Usp.}\ }\textbf
      {\bibinfo {volume} {14}},\ \bibinfo {pages} {673} (\bibinfo {year}
      {1972})}\BibitemShut {NoStop}%
    \bibitem [{\citenamefont {Pomeranchuk}\ and\ \citenamefont
      {Smorodinsky}(1945)}]{RN264}%
      \BibitemOpen
      \bibfield  {author} {\bibinfo {author} {\bibfnamefont {I.}~\bibnamefont
      {Pomeranchuk}}\ and\ \bibinfo {author} {\bibfnamefont {Y.}~\bibnamefont
      {Smorodinsky}},\ }\href@noop {} {\bibfield  {journal} {\bibinfo  {journal}
      {J. Phys. Ussr}\ }\textbf {\bibinfo {volume} {9}},\ \bibinfo {pages} {97}
      (\bibinfo {year} {1945})}\BibitemShut {NoStop}%
    \bibitem [{\citenamefont {Popov}(2001)}]{RN38}%
      \BibitemOpen
      \bibfield  {author} {\bibinfo {author} {\bibfnamefont {V.~S.}\ \bibnamefont
      {Popov}},\ }\href {https://doi.org/Doi 10.1134/1.1358463} {\bibfield
      {journal} {\bibinfo  {journal} {Phys. At. Nucl.}\ }\textbf {\bibinfo {volume}
      {64}},\ \bibinfo {pages} {367} (\bibinfo {year} {2001})}\BibitemShut
      {NoStop}%
    \bibitem [{\citenamefont {Castro~Neto}\ \emph {et~al.}(2009)\citenamefont
      {Castro~Neto}, \citenamefont {Guinea}, \citenamefont {Peres}, \citenamefont
      {Novoselov},\ and\ \citenamefont {Geim}}]{RN296}%
      \BibitemOpen
      \bibfield  {author} {\bibinfo {author} {\bibfnamefont {A.~H.}\ \bibnamefont
      {Castro~Neto}}, \bibinfo {author} {\bibfnamefont {F.}~\bibnamefont {Guinea}},
      \bibinfo {author} {\bibfnamefont {N.~M.~R.}\ \bibnamefont {Peres}}, \bibinfo
      {author} {\bibfnamefont {K.~S.}\ \bibnamefont {Novoselov}},\ and\ \bibinfo
      {author} {\bibfnamefont {A.~K.}\ \bibnamefont {Geim}},\ }\href
      {https://doi.org/10.1103/RevModPhys.81.109} {\bibfield  {journal} {\bibinfo
      {journal} {Rev. Mod. Phys.}\ }\textbf {\bibinfo {volume} {81}},\ \bibinfo
      {pages} {109} (\bibinfo {year} {2009})}\BibitemShut {NoStop}%
    \bibitem [{\citenamefont {Andrei}\ \emph {et~al.}(2012)\citenamefont {Andrei},
      \citenamefont {Li},\ and\ \citenamefont {Du}}]{RN165}%
      \BibitemOpen
      \bibfield  {author} {\bibinfo {author} {\bibfnamefont {E.~Y.}\ \bibnamefont
      {Andrei}}, \bibinfo {author} {\bibfnamefont {G.}~\bibnamefont {Li}},\ and\
      \bibinfo {author} {\bibfnamefont {X.}~\bibnamefont {Du}},\ }\href
      {https://doi.org/10.1088/0034-4885/75/5/056501} {\bibfield  {journal}
      {\bibinfo  {journal} {Rep. Prog. Phys.}\ }\textbf {\bibinfo {volume} {75}},\
      \bibinfo {pages} {056501} (\bibinfo {year} {2012})}\BibitemShut {NoStop}%
    \bibitem [{\citenamefont {Nomura}\ and\ \citenamefont
      {MacDonald}(2006)}]{RN295}%
      \BibitemOpen
      \bibfield  {author} {\bibinfo {author} {\bibfnamefont {K.}~\bibnamefont
      {Nomura}}\ and\ \bibinfo {author} {\bibfnamefont {A.~H.}\ \bibnamefont
      {MacDonald}},\ }\href {https://doi.org/10.1103/PhysRevLett.96.256602}
      {\bibfield  {journal} {\bibinfo  {journal} {Phys. Rev. Lett.}\ }\textbf
      {\bibinfo {volume} {96}},\ \bibinfo {pages} {256602} (\bibinfo {year}
      {2006})}\BibitemShut {NoStop}%
    \bibitem [{\citenamefont {Hwang}\ \emph {et~al.}(2007)\citenamefont {Hwang},
      \citenamefont {Adam},\ and\ \citenamefont {Sarma}}]{RN289}%
      \BibitemOpen
      \bibfield  {author} {\bibinfo {author} {\bibfnamefont {E.~H.}\ \bibnamefont
      {Hwang}}, \bibinfo {author} {\bibfnamefont {S.}~\bibnamefont {Adam}},\ and\
      \bibinfo {author} {\bibfnamefont {S.~Das}\ \bibnamefont {Sarma}},\ }\href
      {https://doi.org/10.1103/PhysRevLett.98.186806} {\bibfield  {journal}
      {\bibinfo  {journal} {Phys. Rev. Lett.}\ }\textbf {\bibinfo {volume} {98}},\
      \bibinfo {pages} {186806} (\bibinfo {year} {2007})}\BibitemShut {NoStop}%
    \bibitem [{\citenamefont {Novikov}(2007)}]{RN291}%
      \BibitemOpen
      \bibfield  {author} {\bibinfo {author} {\bibfnamefont {D.~S.}\ \bibnamefont
      {Novikov}},\ }\href {https://doi.org/10.1103/PhysRevB.76.245435} {\bibfield
      {journal} {\bibinfo  {journal} {Phys. Rev. B}\ }\textbf {\bibinfo {volume}
      {76}},\ \bibinfo {pages} {245435} (\bibinfo {year} {2007})}\BibitemShut
      {NoStop}%
    \bibitem [{\citenamefont {Pereira}\ \emph {et~al.}(2007)\citenamefont
      {Pereira}, \citenamefont {Nilsson},\ and\ \citenamefont
      {Castro~Neto}}]{RN194}%
      \BibitemOpen
      \bibfield  {author} {\bibinfo {author} {\bibfnamefont {V.~M.}\ \bibnamefont
      {Pereira}}, \bibinfo {author} {\bibfnamefont {J.}~\bibnamefont {Nilsson}},\
      and\ \bibinfo {author} {\bibfnamefont {A.~H.}\ \bibnamefont {Castro~Neto}},\
      }\href {https://doi.org/10.1103/PhysRevLett.99.166802} {\bibfield  {journal}
      {\bibinfo  {journal} {Phys. Rev. Lett.}\ }\textbf {\bibinfo {volume} {99}},\
      \bibinfo {pages} {166802} (\bibinfo {year} {2007})}\BibitemShut {NoStop}%
    \bibitem [{\citenamefont {Shytov}\ \emph
      {et~al.}(2007{\natexlab{a}})\citenamefont {Shytov}, \citenamefont
      {Katsnelson},\ and\ \citenamefont {Levitov}}]{RN43}%
      \BibitemOpen
      \bibfield  {author} {\bibinfo {author} {\bibfnamefont {A.~V.}\ \bibnamefont
      {Shytov}}, \bibinfo {author} {\bibfnamefont {M.~I.}\ \bibnamefont
      {Katsnelson}},\ and\ \bibinfo {author} {\bibfnamefont {L.~S.}\ \bibnamefont
      {Levitov}},\ }\href {https://doi.org/10.1103/PhysRevLett.99.246802}
      {\bibfield  {journal} {\bibinfo  {journal} {Phys. Rev. Lett.}\ }\textbf
      {\bibinfo {volume} {99}},\ \bibinfo {pages} {246802} (\bibinfo {year}
      {2007}{\natexlab{a}})}\BibitemShut {NoStop}%
    \bibitem [{\citenamefont {Pereira}\ \emph {et~al.}(2008)\citenamefont
      {Pereira}, \citenamefont {Kotov},\ and\ \citenamefont {Castro~Neto}}]{RN193}%
      \BibitemOpen
      \bibfield  {author} {\bibinfo {author} {\bibfnamefont {V.~M.}\ \bibnamefont
      {Pereira}}, \bibinfo {author} {\bibfnamefont {V.~N.}\ \bibnamefont {Kotov}},\
      and\ \bibinfo {author} {\bibfnamefont {A.~H.}\ \bibnamefont {Castro~Neto}},\
      }\href {https://doi.org/10.1103/PhysRevB.78.085101} {\bibfield  {journal}
      {\bibinfo  {journal} {Phys. Rev. B}\ }\textbf {\bibinfo {volume} {78}},\
      \bibinfo {pages} {085101} (\bibinfo {year} {2008})}\BibitemShut {NoStop}%
    \bibitem [{\citenamefont {Shytov}\ \emph
      {et~al.}(2007{\natexlab{b}})\citenamefont {Shytov}, \citenamefont
      {Katsnelson},\ and\ \citenamefont {Levitov}}]{RN44}%
      \BibitemOpen
      \bibfield  {author} {\bibinfo {author} {\bibfnamefont {A.~V.}\ \bibnamefont
      {Shytov}}, \bibinfo {author} {\bibfnamefont {M.~I.}\ \bibnamefont
      {Katsnelson}},\ and\ \bibinfo {author} {\bibfnamefont {L.~S.}\ \bibnamefont
      {Levitov}},\ }\href {https://doi.org/10.1103/PhysRevLett.99.236801}
      {\bibfield  {journal} {\bibinfo  {journal} {Phys. Rev. Lett.}\ }\textbf
      {\bibinfo {volume} {99}},\ \bibinfo {pages} {236801} (\bibinfo {year}
      {2007}{\natexlab{b}})}\BibitemShut {NoStop}%
    \bibitem [{\citenamefont {Nishida}(2014)}]{RN191}%
      \BibitemOpen
      \bibfield  {author} {\bibinfo {author} {\bibfnamefont {Y.}~\bibnamefont
      {Nishida}},\ }\href {https://doi.org/10.1103/PhysRevB.90.165414} {\bibfield
      {journal} {\bibinfo  {journal} {Phys. Rev. B}\ }\textbf {\bibinfo {volume}
      {90}},\ \bibinfo {pages} {165414} (\bibinfo {year} {2014})}\BibitemShut
      {NoStop}%
    \bibitem [{\citenamefont {Nishida}(2016)}]{RN261}%
      \BibitemOpen
      \bibfield  {author} {\bibinfo {author} {\bibfnamefont {Y.}~\bibnamefont
      {Nishida}},\ }\href {https://doi.org/10.1103/PhysRevB.94.085430} {\bibfield
      {journal} {\bibinfo  {journal} {Phys. Rev. B}\ }\textbf {\bibinfo {volume}
      {94}},\ \bibinfo {pages} {085430} (\bibinfo {year} {2016})}\BibitemShut
      {NoStop}%
    \bibitem [{\citenamefont {Efimov}(1970)}]{RN174}%
      \BibitemOpen
      \bibfield  {author} {\bibinfo {author} {\bibfnamefont {V.}~\bibnamefont
      {Efimov}},\ }\href {https://doi.org/10.1016/0370-2693(70)90349-7} {\bibfield
      {journal} {\bibinfo  {journal} {Phys. Lett. B}\ }\textbf {\bibinfo {volume}
      {B 33}},\ \bibinfo {pages} {563} (\bibinfo {year} {1970})}\BibitemShut
      {NoStop}%
    \bibitem [{\citenamefont {Sornette}(1998)}]{RN247}%
      \BibitemOpen
      \bibfield  {author} {\bibinfo {author} {\bibfnamefont {D.}~\bibnamefont
      {Sornette}},\ }\href {https://doi.org/10.1016/s0370-1573(97)00076-8}
      {\bibfield  {journal} {\bibinfo  {journal} {Phys. Rep.}\ }\textbf {\bibinfo
      {volume} {297}},\ \bibinfo {pages} {239} (\bibinfo {year}
      {1998})}\BibitemShut {NoStop}%
    \bibitem [{\citenamefont {Efimov}(1971)}]{RN175}%
      \BibitemOpen
      \bibfield  {author} {\bibinfo {author} {\bibfnamefont {V.~N.}\ \bibnamefont
      {Efimov}},\ }\href {<Go to ISI>://WOS:A1971J409900025} {\bibfield  {journal}
      {\bibinfo  {journal} {Sov. J. Nucl. Phys.}\ }\textbf {\bibinfo {volume}
      {12}},\ \bibinfo {pages} {589} (\bibinfo {year} {1971})}\BibitemShut
      {NoStop}%
    \bibitem [{\citenamefont {{Liu}}\ \emph {et~al.}(2018)\citenamefont {{Liu}},
      \citenamefont {{Jiang}}, \citenamefont {{Wang}}, \citenamefont {{Joynt}},\
      and\ \citenamefont {{Xie}}}]{RN265}%
      \BibitemOpen
      \bibfield  {author} {\bibinfo {author} {\bibfnamefont {H.}~\bibnamefont
      {{Liu}}}, \bibinfo {author} {\bibfnamefont {H.}~\bibnamefont {{Jiang}}},
      \bibinfo {author} {\bibfnamefont {Z.}~\bibnamefont {{Wang}}}, \bibinfo
      {author} {\bibfnamefont {R.}~\bibnamefont {{Joynt}}},\ and\ \bibinfo {author}
      {\bibfnamefont {X.~C.}\ \bibnamefont {{Xie}}},\ }\href@noop {} {\bibfield
      {journal} {\bibinfo  {journal} {arXiv e-prints}\ ,\ \bibinfo {eid}
      {arXiv:1807.02459}} (\bibinfo {year} {2018})},\ \Eprint
      {https://arxiv.org/abs/1807.02459} {arXiv:1807.02459 [cond-mat.mtrl-sci]}
      \BibitemShut {NoStop}%
    \bibitem [{\citenamefont {Ostrovsky}\ \emph {et~al.}(2006)\citenamefont
      {Ostrovsky}, \citenamefont {Gornyi},\ and\ \citenamefont {Mirlin}}]{RN290}%
      \BibitemOpen
      \bibfield  {author} {\bibinfo {author} {\bibfnamefont {P.~M.}\ \bibnamefont
      {Ostrovsky}}, \bibinfo {author} {\bibfnamefont {I.~V.}\ \bibnamefont
      {Gornyi}},\ and\ \bibinfo {author} {\bibfnamefont {A.~D.}\ \bibnamefont
      {Mirlin}},\ }\href {https://doi.org/10.1103/PhysRevB.74.235443} {\bibfield
      {journal} {\bibinfo  {journal} {Phys. Rev. B}\ }\textbf {\bibinfo {volume}
      {74}},\ \bibinfo {pages} {235443} (\bibinfo {year} {2006})}\BibitemShut
      {NoStop}%
    \bibitem [{\citenamefont {Ovdat}\ \emph {et~al.}(2017)\citenamefont {Ovdat},
      \citenamefont {Mao}, \citenamefont {Jiang}, \citenamefont {Andrei},\ and\
      \citenamefont {Akkermans}}]{RN41}%
      \BibitemOpen
      \bibfield  {author} {\bibinfo {author} {\bibfnamefont {O.}~\bibnamefont
      {Ovdat}}, \bibinfo {author} {\bibfnamefont {J.}~\bibnamefont {Mao}}, \bibinfo
      {author} {\bibfnamefont {Y.}~\bibnamefont {Jiang}}, \bibinfo {author}
      {\bibfnamefont {E.~Y.}\ \bibnamefont {Andrei}},\ and\ \bibinfo {author}
      {\bibfnamefont {E.}~\bibnamefont {Akkermans}},\ }\href
      {https://doi.org/10.1038/s41467-017-00591-8} {\bibfield  {journal} {\bibinfo
      {journal} {Nat. Commun.}\ }\textbf {\bibinfo {volume} {8}},\ \bibinfo {pages}
      {507} (\bibinfo {year} {2017})}\BibitemShut {NoStop}%
    \bibitem [{\citenamefont {Wang}\ \emph {et~al.}(2013)\citenamefont {Wang},
      \citenamefont {Wong}, \citenamefont {Shytov}, \citenamefont {Brar},
      \citenamefont {Choi}, \citenamefont {Wu}, \citenamefont {Tsai}, \citenamefont
      {Regan}, \citenamefont {Zettl}, \citenamefont {Kawakami}, \citenamefont
      {Louie}, \citenamefont {Levitov},\ and\ \citenamefont {Crommie}}]{RN200}%
      \BibitemOpen
      \bibfield  {author} {\bibinfo {author} {\bibfnamefont {Y.}~\bibnamefont
      {Wang}}, \bibinfo {author} {\bibfnamefont {D.}~\bibnamefont {Wong}}, \bibinfo
      {author} {\bibfnamefont {A.~V.}\ \bibnamefont {Shytov}}, \bibinfo {author}
      {\bibfnamefont {V.~W.}\ \bibnamefont {Brar}}, \bibinfo {author}
      {\bibfnamefont {S.}~\bibnamefont {Choi}}, \bibinfo {author} {\bibfnamefont
      {Q.}~\bibnamefont {Wu}}, \bibinfo {author} {\bibfnamefont {H.-Z.}\
      \bibnamefont {Tsai}}, \bibinfo {author} {\bibfnamefont {W.}~\bibnamefont
      {Regan}}, \bibinfo {author} {\bibfnamefont {A.}~\bibnamefont {Zettl}},
      \bibinfo {author} {\bibfnamefont {R.~K.}\ \bibnamefont {Kawakami}}, \bibinfo
      {author} {\bibfnamefont {S.~G.}\ \bibnamefont {Louie}}, \bibinfo {author}
      {\bibfnamefont {L.~S.}\ \bibnamefont {Levitov}},\ and\ \bibinfo {author}
      {\bibfnamefont {M.~F.}\ \bibnamefont {Crommie}},\ }\href
      {https://doi.org/10.1126/science.1234320} {\bibfield  {journal} {\bibinfo
      {journal} {Science}\ }\textbf {\bibinfo {volume} {340}},\ \bibinfo {pages}
      {734} (\bibinfo {year} {2013})}\BibitemShut {NoStop}%
    \bibitem [{\citenamefont {Wang}\ \emph {et~al.}(2018)\citenamefont {Wang},
      \citenamefont {Liu}, \citenamefont {Li}, \citenamefont {Liu}, \citenamefont
      {Wang}, \citenamefont {Liu}, \citenamefont {Dai}, \citenamefont {Wang},
      \citenamefont {Li}, \citenamefont {Yan}, \citenamefont {Mandrus},
      \citenamefont {Xie},\ and\ \citenamefont {Wang}}]{RN36}%
      \BibitemOpen
      \bibfield  {author} {\bibinfo {author} {\bibfnamefont {H.}~\bibnamefont
      {Wang}}, \bibinfo {author} {\bibfnamefont {H.}~\bibnamefont {Liu}}, \bibinfo
      {author} {\bibfnamefont {Y.}~\bibnamefont {Li}}, \bibinfo {author}
      {\bibfnamefont {Y.}~\bibnamefont {Liu}}, \bibinfo {author} {\bibfnamefont
      {J.}~\bibnamefont {Wang}}, \bibinfo {author} {\bibfnamefont {J.}~\bibnamefont
      {Liu}}, \bibinfo {author} {\bibfnamefont {J.-Y.}\ \bibnamefont {Dai}},
      \bibinfo {author} {\bibfnamefont {Y.}~\bibnamefont {Wang}}, \bibinfo {author}
      {\bibfnamefont {L.}~\bibnamefont {Li}}, \bibinfo {author} {\bibfnamefont
      {J.}~\bibnamefont {Yan}}, \bibinfo {author} {\bibfnamefont {D.}~\bibnamefont
      {Mandrus}}, \bibinfo {author} {\bibfnamefont {X.~C.}\ \bibnamefont {Xie}},\
      and\ \bibinfo {author} {\bibfnamefont {J.}~\bibnamefont {Wang}},\ }\href
      {https://doi.org/10.1126/sciadv.aau5096} {\bibfield  {journal} {\bibinfo
      {journal} {Sci. Adv.}\ }\textbf {\bibinfo {volume} {4}},\ \bibinfo {pages}
      {eaau5096} (\bibinfo {year} {2018})}\BibitemShut {NoStop}%
    \bibitem [{\citenamefont {Dresselhaus}\ and\ \citenamefont
      {Dresselhaus}(2002)}]{RN292}%
      \BibitemOpen
      \bibfield  {author} {\bibinfo {author} {\bibfnamefont {M.~S.}\ \bibnamefont
      {Dresselhaus}}\ and\ \bibinfo {author} {\bibfnamefont {G.}~\bibnamefont
      {Dresselhaus}},\ }\href {https://doi.org/10.1080/00018730110113644}
      {\bibfield  {journal} {\bibinfo  {journal} {Adv. Phys.}\ }\textbf {\bibinfo
      {volume} {51}},\ \bibinfo {pages} {1} (\bibinfo {year} {2002})}\BibitemShut
      {NoStop}%
    \bibitem [{\citenamefont {Mao}\ \emph {et~al.}(2016)\citenamefont {Mao},
      \citenamefont {Jiang}, \citenamefont {Moldovan}, \citenamefont {Li},
      \citenamefont {Watanabe}, \citenamefont {Taniguchi}, \citenamefont {Masir},
      \citenamefont {Peeters},\ and\ \citenamefont {Andrei}}]{RN187}%
      \BibitemOpen
      \bibfield  {author} {\bibinfo {author} {\bibfnamefont {J.}~\bibnamefont
      {Mao}}, \bibinfo {author} {\bibfnamefont {Y.}~\bibnamefont {Jiang}}, \bibinfo
      {author} {\bibfnamefont {D.}~\bibnamefont {Moldovan}}, \bibinfo {author}
      {\bibfnamefont {G.}~\bibnamefont {Li}}, \bibinfo {author} {\bibfnamefont
      {K.}~\bibnamefont {Watanabe}}, \bibinfo {author} {\bibfnamefont
      {T.}~\bibnamefont {Taniguchi}}, \bibinfo {author} {\bibfnamefont {M.~R.}\
      \bibnamefont {Masir}}, \bibinfo {author} {\bibfnamefont {F.~M.}\ \bibnamefont
      {Peeters}},\ and\ \bibinfo {author} {\bibfnamefont {E.~Y.}\ \bibnamefont
      {Andrei}},\ }\href {https://doi.org/10.1038/nphys3665} {\bibfield  {journal}
      {\bibinfo  {journal} {Nat. Phys.}\ }\textbf {\bibinfo {volume} {12}},\
      \bibinfo {pages} {545} (\bibinfo {year} {2016})}\BibitemShut {NoStop}%
    \bibitem [{\citenamefont {Wang}\ \emph {et~al.}(2012)\citenamefont {Wang},
      \citenamefont {Brar}, \citenamefont {Shytov}, \citenamefont {Wu},
      \citenamefont {Regan}, \citenamefont {Tsai}, \citenamefont {Zettl},
      \citenamefont {Levitov},\ and\ \citenamefont {Crommie}}]{RN298}%
      \BibitemOpen
      \bibfield  {author} {\bibinfo {author} {\bibfnamefont {Y.}~\bibnamefont
      {Wang}}, \bibinfo {author} {\bibfnamefont {V.~W.}\ \bibnamefont {Brar}},
      \bibinfo {author} {\bibfnamefont {A.~V.}\ \bibnamefont {Shytov}}, \bibinfo
      {author} {\bibfnamefont {Q.}~\bibnamefont {Wu}}, \bibinfo {author}
      {\bibfnamefont {W.}~\bibnamefont {Regan}}, \bibinfo {author} {\bibfnamefont
      {H.-Z.}\ \bibnamefont {Tsai}}, \bibinfo {author} {\bibfnamefont
      {A.}~\bibnamefont {Zettl}}, \bibinfo {author} {\bibfnamefont {L.~S.}\
      \bibnamefont {Levitov}},\ and\ \bibinfo {author} {\bibfnamefont {M.~F.}\
      \bibnamefont {Crommie}},\ }\href {https://doi.org/10.1038/nphys2379
      https://www.nature.com/articles/nphys2379#supplementary-information}
      {\bibfield  {journal} {\bibinfo  {journal} {Nat. Phys.}\ }\textbf {\bibinfo
      {volume} {8}},\ \bibinfo {pages} {653} (\bibinfo {year} {2012})}\BibitemShut
      {NoStop}%
    \bibitem [{\citenamefont {Landau}\ and\ \citenamefont
      {Lifshitz}(2013)}]{RN259}%
      \BibitemOpen
      \bibfield  {author} {\bibinfo {author} {\bibfnamefont {L.~D.}\ \bibnamefont
      {Landau}}\ and\ \bibinfo {author} {\bibfnamefont {E.~M.}\ \bibnamefont
      {Lifshitz}},\ }\href@noop {} {\emph {\bibinfo {title} {Quantum mechanics:
      non-relativistic theory}}},\ Vol.~\bibinfo {volume} {3}\ (\bibinfo
      {publisher} {Elsevier},\ \bibinfo {year} {2013})\BibitemShut {NoStop}%
    \bibitem [{\citenamefont {Silvestrov}\ and\ \citenamefont
      {Efetov}(2007)}]{RN293}%
      \BibitemOpen
      \bibfield  {author} {\bibinfo {author} {\bibfnamefont {P.~G.}\ \bibnamefont
      {Silvestrov}}\ and\ \bibinfo {author} {\bibfnamefont {K.~B.}\ \bibnamefont
      {Efetov}},\ }\href {https://doi.org/10.1103/PhysRevLett.98.016802} {\bibfield
       {journal} {\bibinfo  {journal} {Phys. Rev. Lett.}\ }\textbf {\bibinfo
      {volume} {98}},\ \bibinfo {pages} {016802} (\bibinfo {year}
      {2007})}\BibitemShut {NoStop}%
    \bibitem [{\citenamefont {Chen}\ \emph {et~al.}(2007)\citenamefont {Chen},
      \citenamefont {Apalkov},\ and\ \citenamefont {Chakraborty}}]{RN294}%
      \BibitemOpen
      \bibfield  {author} {\bibinfo {author} {\bibfnamefont {H.-Y.}\ \bibnamefont
      {Chen}}, \bibinfo {author} {\bibfnamefont {V.}~\bibnamefont {Apalkov}},\ and\
      \bibinfo {author} {\bibfnamefont {T.}~\bibnamefont {Chakraborty}},\ }\href
      {https://doi.org/10.1103/PhysRevLett.98.186803} {\bibfield  {journal}
      {\bibinfo  {journal} {Phys. Rev. Lett.}\ }\textbf {\bibinfo {volume} {98}},\
      \bibinfo {pages} {186803} (\bibinfo {year} {2007})}\BibitemShut {NoStop}%
    \end{thebibliography}
%
    
\end{document}